\documentstyle[12pt,a4,epsfig]{article}

\newcommand{\beq}{\begin{equation}}
\newcommand{\eeq}{\end{equation}}
\newcommand{\beqar}[1]{\begin{eqnarray}\label{#1}}
\newcommand{\eeqar}{\end{eqnarray}}

\newcommand{\ga}{\gamma}
\newcommand{\de}{\delta}   
\newcommand{\De}{\Delta}

\newcommand{\ph}{\varphi} 
\newcommand{\si}{\sigma}

\newcommand{\om}{\omega}

\newcommand{\La}{\Lambda}

\newcommand{\as}{\alpha_S}

\def\eq#1{{Eq.~(\ref{#1})}} 
\def\npb#1#2#3{    {\it Nucl. Phys. }{\bf B#1} (19#2) #3}
\def\plb#1#2#3{    {\it Phys. Lett. }{\bf B#1} (19#2) #3}
\def\prd#1#2#3{    {\it Phys. Rev. }{\bf D#1} (19#2) #3}

\relax
\begin{document}
\vspace*{-2cm}
\begin{flushright}
TAUP 2501-98\\
{\tt hep - ph/9806228}\\
June 2, 1998
\end{flushright}
\vspace{2cm}
\begin{center}
{\LARGE  \bf
 The BFKL high energy  asymptotics\\ in the next-to-leading approximation}
\vspace{1cm}

{\Large \tt   Eugene  Levin ${}^{\ast}$}\\[1.5ex]
 
{\large \tt  $^{\ast}$ Email: leving@post.tau.ac.il.}\\[1.0ex]

{\it  School of Physics and Astronomy}\\

{\it Raymond and Beverly Sackler Faculty of Exact Science}\\

{\it Tel Aviv University, Tel Aviv, 69978, ISRAEL}\\[1.5ex]

\end{center}
\vspace{4cm}
\centerline{\bf Abstract:}
We discuss the high energy asymptotics in the next-to-leading (NLO) BFKL
equation. We find a general solution for Green functions and consider two
properties  of the NLO BFKL kernel: running QCD coupling and large NLO
corrections to the conformal part of the kernel. Both of these effects 
lead to Regge-BFKL asymptotics only in the limited range of energy ( $y =
\ln(s/q q_0)\,\leq\,(\as)^{ - \frac{5}{3}}$ ) and change the energy
behaviour of the amplitude for higher values of energy.  We confirm the
oscillation in the total cross section  found in Ref. \cite{ROSS} in the
NLO BFKL asymptotics, which shows that the NLO BFKL has a  serious
pathology.

\newpage

\section{Introduction }
The next-to-leading order corrections to the kernel of the BFKL
equation \cite{BFKL} (NLO BFKL)  have recently  been presented  in two
papers
\cite{CC}\cite{FL}, and 
have been formulated in a compact and elegant form by Fadin and Lipatov
\cite{FL}. It turns out that the next-to-leading order corrections are
large and they strongly modify  the leading order result.  The large
value
of the next-to-leading order corrections suggests that even high order
corrections may  be  essential, leaving us without   a reliable
calculation
of the parameters of the BFKL asymptotics except at very small
values of the QCD coupling constant, which unfortunately are not presently
attainable.

 Before arriving at  a
pessimistic conclusion, it is necessary   to understand the 
qualitative alteration due to the
next-to-leading order corrections that occurs in the BFKL asymptotics,
i.e. we wish to know  how the NLO BFKL Pomeron differs from the LO BFKL
one.  Other questions   which we would like to answer are:
Does  the NLO BFKL Pomeron still 
 manifest itself as a Reggeon - like exchange at high energy as the LO
BFKL Pomeron does? Can  the NLO BFKL Pomeron be  described as the
diffusion cascading process in log $k_{\perp}$ where $k_{\perp} $ is the
partonic transverse momentum? Can we  calculate the NLO BFKL Pomeron in
the framework of perturbative QCD or the nonperturbative correction
would destroy   our approach at so small value of energy that we cannot
use the BFKKL asymptotic? In short, we  would like  to understand what 
the NLO
BFKL Pomeron is ?

This  paper is an attempt to find a general solution to the NLO BFKL
equation,  and
to discuss the main properties of the high energy asymptotics that follows
from this solution.  The NLO BFKL kernel contains two parts: a
conformally invariant part and a running coupling part. In studying  the
running coupling effect we reproduce    non-Regge  type corrections 
to high energy asymptotics suggested by Kovchegov and Mueller in
Ref.\cite{KM}, using a method of solution proposed  in
Refs.\cite{GLR}\cite{LREN}. Ross found in Ref.\cite{ROSS} that the
conformal part of the  NLO BFKL kernel crucially changes  the
diffusion
in log $k_{\perp}$  ($k_{\perp} $ is a  parton transverse momentum  )
which is  a basic property of the LO BFKL Pomeron. We   show how this  
affects 
the asymptotic behaviour  of the scattering amplitude at high energy.

The running QCD coupling in the BFKL equation has been studied for
sometime,
starting with  the GLR paper \cite{GLR}, where the general solution to
this
problem was suggested. In Refs. \cite{LI86}\cite{ORSAY}\cite{HR} \cite{NZ}
\cite{LREN}\cite{BRAUN}\cite{HKK}\cite{KM} the different aspects of the
problem were  considered, but only now this can be done  on
solid basis of
the NLO BFKL equation. The influence of the NLO on the BFKL diffusion was
first
 pointed out and explored  in Ref. \cite{ROSS}.

Our paper  deals with both problems on the same footing, and  we give a
simple and transparent derivation and discussion of the
high energy asymptotics for the NLO BFKL equation. Our conclusion is
rather
pessimistic: we cannot avoid an oscillation in the total cross section at
high energy in the NLO BFKL approach and, therefore, we do not think that
the NLO BFKL equation can serve as a basis for high energy phenomenology.
However, we hope, that the experience with the NLO BFKL Pomeron will be
useful for  possible scenario of the high energy behaviour in QCD.

\section{A general approach to the  NLO BFKL equation}
\setcounter{equation}{0}

The BFKL equation governs the high energy asymptotics of a
single-scale hard process. As an example of such a process we can consider
the scattering of two virtual photons ( with virtualities $Q^2$ and
$Q^2_0$ ) at high energy in the kinematic region where $ Q^2
\,\cong\,Q^2_0 \,\,\gg\,m^2$ \cite{PHPH},where $m^2$ a scale of the
``soft" interaction. The cross section for this
process can be
written as
\beq \label{1.1}
\si_{\ga^* \ga^*} (s)\,\,=
\eeq
$$
\int\,\frac{d^2 q}{2 \pi q^2}\,\int\,
 \frac{d^2 q_0}{2 \pi q^2_0}\,\Phi(Q^2,q^2)\,\Phi(Q^2_0,q^2_0)\,
\int^{a + i\infty}_{a - i\infty}\,\frac{d \om}{2 \pi
i}\,\left(\,\frac{s}{q\,q_0}
\right)^{\om}\,G_{\om}(\vec{q},\vec{q}_0)\,\,,
$$
where $\vec{q}$ and $\vec{q}_0$ are transverse momenta of gluons, $\Phi
(Q^2,q^2)$
and $\Phi(Q^2_0,q^2_0)$ are impact functions that has been calculated in
Ref.\cite{PHPH}. It should be stressed that functions
$\Phi$ provide $ q^2 \,\cong\, Q^2$ and $q^2_0\,\cong\,
Q^2_0$. $G_{\om}(\vec{q},\vec{q}_0)$ obeys the BFKL equation \beq
\label{BFKL}
\om\,G_{\om}(\vec{q},\vec{q}_0)\,\,=\,\,\delta^{(2)}( \vec{q} -  \vec{q}_0
)\,\,+\,\,\int\,d^2 
q'\,K(\vec{q},\vec{q} ')\,G_{\om}(\vec{q} ',\vec{q}_0)\,\,.
\eeq
Kernel $K(\vec{q},\vec{q} ')$ can be written as a sum of LO and NLO
kernels
\beq \label{1.3}
K(\vec{q},\vec{q} ')\,\,=\,\,K^{LO}(\vec{q},\vec{q} ')\,\,+
\,\,K^{NLO}(\vec{q},\vec{q} ')\,\,.
\eeq
This kernel has eigenfunction\cite{FL} 
\beq \label{1.4}
\ph_f(q^2)\,\,=\,\,\frac{1}{\sqrt{q^2}}\,
\left(\,\frac{q^2}{\sqrt{\as(q^2)}}\,\right)^{f}
\eeq
and corresponding eigenvalues
\beq \label{1.5}
\int\,d^2 q'
K(\vec{q},\vec{q} ')\,\ph_f(q'^2)\,\,=\,\,\om(f)\,\ph_f(q^2)\,\,.
\eeq
 $\om(f)$ has a form
\beq \label{OMEGA}
\om(f)\,\,=
\eeq
$$
\bar \as( \mu^2 )\,(\,\chi^{LO}(f)\,\,+\,\,\bar
\as (\mu^2)\,\chi^{NLO}(f)\,)\,\,-\,\, \frac{ N_c
\,\as^2(\mu^2)}{\pi}\,b\,\ln (q^2/\mu^2)\,\chi^{LO}(f)\,\,,
$$
where $\chi^{LO}(f)\,\,=\,\,3\psi(1)\,-\,\psi(\frac{1}{2} +f)\,
-\,\psi(\frac{1}{2} - f)$,  the explicit form of $\chi^{NLO}$ is written
in Ref. \cite{FL}, which has also symmetry under transform $f
\,\rightarrow\, - f$ as the LO part, and where $b\,\,=\,\,\frac{11 N_c -
2N_f}{12 \,\pi}$ for number of colours $N_c$ and number of flavours $N_f$.
We use notation $\bar \as$ for $\bar \as = \frac{N_c \as}{\pi}$.

One can see that the first term has a conformal symmetry while this
symmetry is broken in the NLO due to the second term in \eq{OMEGA}.
$\mu^2$ is the normalization point which is arbitrary in the NLO
calculations. We take  it to be  equal to the value of the  initial
virtuality
$q^2_0$ ( $\mu^2\,=\,q^2_0$ ) without loosing any accuracy in the NLO
approach.

Following Refs. \cite{GLR} \cite{LREN} we rewrite \eq{OMEGA} in the form:
\beq \label{OMEGA1}
\om(f)\,\,=\,\,\frac{r_0}{r}\,\bar \as( q^2_0)\,(\,\chi^{LO}(f)\,\,+\,\,\bar
\as (q^2_0)\,\chi^{NLO}(f)\,)\,\,=\,\,\frac{r_0}{r}\,\om_{conf}(f)\,\,.
\eeq
Here we define $ r\,\,=\,\,\ln\frac{q^2}{\La^2}\,-\,\frac{1}{2}\ln\as(q^2)$ 
and  the running QCD coupling constant in leading log   is equal
to \footnote{We add factor $\frac{1}{2}\ln\as(q^2)$ for 
convenience  but it
does not affect the value of $\om (f)$ in the NLO approximation.} 
$$
\as(q^2)\,\,=\,\,\frac{1}{b \,r}\,\,,
$$ 
where $\La^2$ is the position of the infrared Landau pole in running
$\as$.

We want to stress that \eq{OMEGA1} coincides with \eq{OMEGA} in the NLO
approximation. However, this form of the kernel is much more convenient in
searching of a solution and we firmly believe it corresponds more to  the
general incorporation of the effect of the running QCD coupling in the
BFKL equation ( see a  discussion of this point of view in Ref.
\cite{LREN} ).

To find a solution to \eq{BFKL} we expand $ G_{\om}( q^2) $ with respect
to the complete set of eigenfunction  of \eq{1.4}, namely
\beq \label{MELLIN}
G_{\om}(q^2)\,\,=
\eeq
$$
\,\,\int^{a \,+\,i\infty}_{a - i\infty}\,\frac{d f}{2 \pi
i}\,\,g( \om,f)\,\,
\ph_f(q^2)\,\,=\,\,\frac{1}{\sqrt{q^2\,q^2_0}}\,\,\int^{a
\,+\,i\infty}_{a -i\infty}\,
\frac{d f}{2 \pi i}\,\,g( \om, f)\,\,e^{r f} \,\,,
$$
where the contour of integration is situated to the right of  all
singularities of function $g(\om,f)$.

Using  \eq{1.5} and \eq{OMEGA1} one obtains  the
following equation for function $g( \om, f)$ :
\beq \label{RUNBFKL}
-\,\om \frac{d g (\om, f)}{d f}\,\,=\,\,r_0\,\om_{conf} (f) \,g (\om,
f)\,\,+\,\,r_0 \,e^{-f r_0}\,\,.
\eeq
The solution of homogeneous equation ( \eq{RUNBFKL} without the last term
) can be easily found and it has the  form ( see Refs. \cite{GLR}
\cite{LREN}
for details ):
\beq \label{SOL}
g(\om, f)\,\,=\,\,\tilde g(\om) \,e^{-
\,\frac{r_0}{\om}\,\int^f_{f_0}\,\om_{conf} (f') d f'}\,\,.
\eeq
Function $\tilde g (\om)$ should be specified from initial or boundary
conditions.   The value of $f_0$ can be arbitrary since its redefinition
is included in function $\tilde g(\om)$. Unless  it is  specially
stipulated  $f_0$ = 0.

We will show that 
the small values of $f$ will be dominant in \eq{MELLIN}in the wide range
of large $y \,\,=\,\,\ln\frac{s}{\sqrt{q^2\,q^2_0}}$.  Therefore, we can
expand $\om_{conf}$ at small $f$, namely,
\beq \label{OMD}
 \om_{conf}\,\,=\,\,\om_L\,\,+\,\,D\,f^2\,\,+\,\,O(f^4)\,\,,
\eeq
where in the NLO $\om_L$ and $D$ are equal to ( see Eq.(7) of
Ref.\cite{ROSS} ):
\begin{eqnarray} \label{PAR}
&
\om_L\,\,=\,\,\bar \as(q^2_0)\,\{\, 2.772 \,\,-\,\,18.3 \bar
\as(q^2_0)\,\}\,\,; &\\
&
D\,\,=\,\,\bar \as(q^2_0)\,\{\, 16.828 \,\,-\,\,322 \bar
\as(q^2_0)\,\}\,\,.&
\end{eqnarray}
 
One can see from \eq{PAR} how large and essential the NLO corrections
are.
They considerably diminish the value of $\om_L$ which can be even negative
for $\bar \as(q^2_0) \,>\,0.152$ and change the sign of $D$ at $\bar \as
\approx 0.05$. Note, that the positive value of $D$ corresponds to
diffusion in $\ln  q^2$.

To solve this  problem we need to find $\tilde g(\om)$ in the general
solution of \eq{SOL}, which depends on the initial or boundary conditions.
We find it very instructive to introduce two Green  functions for the BFKL
equation. 

{\bf 1.}

The first one  ( $G_r (y,r)$ ) satisfies the following boundary
condition:
\begin{eqnarray} \label{GREENR}
&
G_r (y,r):&  \\
&
G_r(y, r = r_0)\,\,=\,\,\delta(y \,-\,y_0) &\nonumber
\end{eqnarray}
This Green function allows us to find us the solution of the BFKL equation
for any boundary input  distribution  $G_{in}(y,q^2=q^2_0)$
 at $q^2 = q^2_0$ ( $r = r_0$ ). Indeed, such a solution is equal to
\beq \label{SOLR}
G(y,r)\,\,=\,\,\int \,d y_0 \,G_r(y, r) \,G_{in}(y_0,q^2=q^2_0)
\,\,.
\eeq
Such a  Green function is very useful for study of the boundary
condition for the DGLAP evolution. Using $G_r(y,r)$ and \eq{SOLR},  we can     
investigate the $y$-dependence at $q^2\,\cong\,q^2_0$. We can distinguish
two cases with different solutions:

1. the integral over $y_0$ depends mostly on properties of input function
$G_{in}$;

2. the integral over $y_0$ is sensitive to the Green function. In this
case we can claim that the energy behaviour of our boundary condition is
defined by the BFKL dynamics.

Therefore, this Green function ( $G_r (y,r)$ ) can provide us with  an
educated
guess  for the energy dependence of the boundary condition in the DGLAP
evolution equations \cite{DGLAP}.

In the LO BFKL approach $G_r(y - y_0,r,r_0)$ is equal to\cite{BFKL}
\beq \label{LOGR}
G^{LO}_r(y - y_0,r,r_0)\,\,=\,\,\sqrt{\frac{\pi (r - r_0)}{2 D^{LO} (y -
y_0)^3}}\,\,
e^{\om^{LO}_L\,( y - y_0 )\,\,-\,\,\frac{( r -  r_0 )^2}{4 D^{LO} ( y -
y_0
)}}\,\,.
\eeq
We call  this expression  Regge-BFKL asymptotics since it has a
power - like behaviour, similar to the exchange of the Reggeon.
We
would like to recall that the second factor in the exponent is different
from a  Reggeon contribution, and it has an origin in the diffusion in the
log of transverse momentum of partons which is a typical feature of QCD.

The question is how general is  \eq{LOGR} and does  it preserve the main
characteristics like  power behaviour and/or the diffusion in the log of
transverse momentum in the NLO approximation.

{\bf 2.}

 The initial condition for the second Green function ( $G_y (y,r)$ ) 
can be written as follows:
\begin{eqnarray} \label{GREENY}
&
G_y (y,r):&  \\
&
G_y(y = y_0, r )\,\,=\,\,\delta(r \,-\,r_0) &\nonumber
\end{eqnarray}

It is obvious that one can find a general solution using this Green
function if we have an input from experiment and/or nonperturbative QCD,
namely, the dependence on transverse momentum at fixed value of $y$.
Indeed,
\beq \label{SOLY}
G(y,r)\,\,=\,\,\int \,d r_0 \,G_y(y, r) \,G_{in}(y=y_0,r_0)
\,\,.
\eeq
From \eq{SOLY} one can see that $G_y(y,r)$ defines the asymptotic of the
scattering amplitude for a single-scale process. For example, this Green
function gives  the asymptotic behaviour of $\sigma_{\ga^* \ga^*}$
\cite{PHPH}.

The last remark in this general section: it is necessary to choose
sufficiently
large initial transverse momentum ( $q^2_0$ ) in order  to safely apply
 the pQCD
methods. It has been discussed in many
papers\cite{LI86}\cite{BAR}\cite{MU97}\cite{LL} \cite{HKK}\cite{KM} that
at high
energies  the BFKL diffusion in log of transverse momenta  inevitabelly
leads to the fact that small values of the transverse momenta become
important (Bartel's cigar \cite{BAR}). It means that we cannot
safely calculate
 the high energy
asymptotics in the framework of pQCD.  The criteria for being
able to trust pQCD formulated in
Ref.\cite{MU97} (see also Ref. \cite{KM} for discussion in the case of
running QCD  coupling ) is 
\beq \label{CRTRA}
y\,\,\leq\,\frac{\pi}{14\,N_c\,\zeta(3)\,b^2}\,\times\,\frac{1}{\as^3(q^2_0)}
\eeq
which suggests that we should take $q^2_0$ as large as possible

In the LO BFKL approach this Green function has a similar form as $G_r$,
namely \cite{BFKL}
\beq \label{LOGY}
G^{LO}_y(y - y_0,r,r_0)\,\,=\,\,\sqrt{\frac{\pi}{2 D^{LO} (y -
y_0)}}\,\,
e^{\om^{LO}_L\,( y - y_0 )\,\,-\,\,\frac{( r -  r_0 )^2}{4 D^{LO} ( y -
y_0
)}}\,\,.
\eeq
One can see that both Green functions in the LO BFKL approach are 
similar  and only differ  in pre-exponential factors.

\section{$\mathbf G_r(y,r)$}
\setcounter{equation}{0}
\subsection{Solution}

This Green function has been calculated in Ref.\cite{LREN} but for
completeness we will reproduce simple calculation to examine what
happens    to  $G_r$ in the NLO. Substituting \eq{OMD} in \eq{SOL} we find
that the general solution of the homogeneous equation is
\beq \label{3.1}
G(y,r)\,\,=\,\,\int^{a + i\infty}_{a - i \infty} \,\frac{d \om}{
2 \,\pi\,i}\int^{f_0 + i \infty}_{f_0 - i \infty}\,\frac{d f}{2\,\pi\,i}
\tilde g(\om)\,e^{ \om \,(y - y_0) \,+\,f\,r
\,-\,\frac{r_0\,(\,\om_L f \,+\frac{D}{3}\,f^3\,)}{\om}}\,\,.
\eeq
The integration over $f$ leads to Airy function \cite{LREN}  $  
Ai\left( (\frac{\om}{r_0
D})^{\frac{1}{3}}\,[\,r\,-\,\frac{\om_L}{\om}r_0\,]\,\right)$. Therefore
 to satisfy the boundary condition of \eq{GREENR} we have to choose a
function $\tilde g(\om) \,=\,Ai^{-1}\left( (\frac{\om}{r_0
D})^{\frac{1}{3}}\,[\,r_0\,-\,\frac{\om_L}{\om}r_0\,]\,\right)$.

Finally\cite{LREN}, $G_r(y-y_0,r,r_0)$ is equal to
\beq \label{GRNRF}
G_r(y-y_0,r,r_0)\,\,=\,\,\sqrt{\frac{r}{r_0}}\,\int^{a + i\infty}_{a -
i\infty}\,\frac{d \om}{2\pi i}\,e^{\om(y - y_0)}\,\,\frac{Ai\left(
(\frac{\om}{r_0
D})^{\frac{1}{3}}\,[\,r\,-\,\frac{\om_L}{\om}r_0\,]\,\right)}{Ai\left(
(\frac{\om}{r_0
D})^{\frac{1}{3}}\,[\,r_0\,-\,\frac{\om_L}{\om}r_0\,]\,\right)}\,\,.
\eeq

\subsection{Regge-BFKL asymptotics}
 To recover the Regge-BFKL asymptotics we assume that the arguments of
both Airy functions in \eq{GRNRF}are large:
\begin{eqnarray} \label{ARG}
&
(\frac{\om}{r_0
D})^{\frac{1}{3}}\,[\,r\,-\,\frac{\om_L}{\om}r_0\,]=(\frac{\om}{r_0
D})^{\frac{1}{3}}\,r_0[\,\frac{\de
r}{r_0}\,-\,\frac{\De}{\om_L}\,]\,\,\gg\,1\,\,;& \nonumber\\
&
(\frac{\om}{r_0
D})^{\frac{1}{3}}\,[\,r_0\,-\,\frac{\om_L}{\om}r_0\,]=(\frac{\om}{r_0
D})^{\frac{1}{3}}\,r_0[\,\frac{\De}{\om_L}\,]\,\,\gg\,1\,\,;&  
\end{eqnarray}
where $\de r\,=\,r\,-\,r_0$ and $\De \,=\,\om\,-\,\om_L$.

Using the asymptotics of Airy function \cite{MATH}
$Ai(z)|_{z\,>\,0 ; |z|\,\gg\,1}\,\,\rightarrow\,\,\frac{1}{2
z^{\frac{1}{4}}}
e^{-\frac{2}{3} \,z^{\frac{3}{2}}}$  we obtain for positive $D$
($D\,>\,0$): 
\beq \label{ASYMR}
G_r(y-y_0,r,r_0)\,\,=\,\,\int^{a + i\infty}_{a - i \infty}\,\frac{d \om}{2
\pi i}\,e^{\Psi(\Delta,y,r)}\,\,,
\eeq
where $\Psi$ is equal to
\beq \label{PSI}
\Psi\,=\,\om_L (y - y_0) \,+\,\De (y - y_0) \,-\,
\frac{2}{3}\,\sqrt{\frac{\om_L}{D}}\,r_0\,\,\left(\,
\{ \,\frac{\de r}{r_0}\,\,+\,\,\frac{\De}{\om_L}\,\}^{\frac{3}{2}}
\,\,-\,\,\{ \frac{\De}{\om_L}\,\}^{\frac{3}{2}}\,\right)\,\,.
\eeq
We can calculate this integral using the saddle point method, in which the
value of saddle point ($\De^S$) is defined from the equation:
\beq \label{SP}
\frac{d \Psi}{d \De}\,|_{\De = \De^S}\,\,=\,\,0\,\,.
\eeq 
 In our case this equation reads
\beq \label{SP1}
 (y - y_0) \,-\,
\,\sqrt{\frac{1}{D\,\om_L}}\,r_0\,\,\left(\,
\{ \,\frac{\de
r}{r_0}\,\,+\,\,\frac{\De^S}{\om_L}\,\}^{\frac{1}{2}}
\,\,-\,\,\{ \frac{\De^S}{\om_L}\,\}^{\frac{1}{2}}\,\right)\,\,=\,\,0
\eeq
We can easily find a solution to \eq{SP1}, assuming that
\beq \label{SP2}
\kappa\,\,=\,\,\frac{\de r\,\,\om_L}{r_0\,\,\De}\,\,\ll\,\,1\,\,.
\eeq
\eq{SP1} can be reduced to the form
\beq \label{SP01}
\sqrt{\De^S}\,\,=\,\,\frac{\de r}{2\,\sqrt{D}\,( y - y_0
)}\,\left(\,\,1\,\,-\,\,\frac{1}{4}\,\kappa^S\,\,-\,\,\frac{1}{8}\,
(\,\kappa^S\,)^2\,\,+\,\,O\left(( \kappa^S )^3 \right))\,\,\right)
\eeq

In the leading order with respect to $\kappa^S$,  the solution to
\eq{SP01}
has  the form:
\beq \label{SP3}
\sqrt{\De^S_0}\,\,=\,\,\frac{\de r}{ 2 \sqrt{D} (y - y_0)}\,\,;
\,\,\,\,\,\,\De^S_0\,\,=\,\,\frac{ (\de r)^2}{4 D (y - y_0)^2}\,\,.
\eeq
However, we need to calculate the value of the saddle point with better
accuracy to calculate  the deviation from the Regge-BFKL behaviour. 
Considering  
 $\kappa^{S}_0\,\,=\,\,\frac{ 4 D \om_L (y - y_0)^2}{r_0\,\de r} $ 
 being small, we find
\beq \label{SP4}
\sqrt{\De^S}\,\,=\,\,\,\frac{\de r}{ 2 \sqrt{D} (y - y_0)}\,\left(\,
1\,\,-\,\,\frac{1}{4}\,\kappa^S_0\,\,+\,\,
O\left( (\kappa^S_0)^3\right)\,\right)\,\,.
\eeq
The integrand (see \eq{PSI} ) for $\De \,=\,\De^S$ is equal to 
\beq \label{PSISP}
\Psi(\De^S) = \om_L (y - y_0)\,(\,1 - \frac{\de r}{2
r_0}\,)\,\,-\,\,\frac{(\de
r)^2}{4 D (y - y_0)}\,\,+\,\, \frac{1}{12}
\,\frac{D \om^2_L}{r^2_0}\,(y - y_0)^3\,\,.
\eeq
 One can see that for positive $D$ we have a good saddle point, and can
evaluate the  integral using the steepest descent method (SDM ). The
result is
\beq \label{SPA}
G_r(y - y_0,r)\,\,=\,\,\frac{1}{q q_0} \sqrt{\frac{\pi (\de r)^2}{ 2 D (y
- y_0)^3} }\, 
e^{\Psi(\De^S)}\,\,.   
\eeq
 
The first three terms in \eq{PSISP} yield  normal Regge-BFKL asymptotic
form
which is the same as in the LO BFKL approach. The difference is only in
the second term which fixes the scale of the running QCD coupling in the 
expression for $\om_L$ (see \eq{PAR} ), namely in the BFKL diffusion, the
running QCD coupling enters at the scale $q^2_{av}\,\,=\,\,q q_0)$.
In other words we have  to calculate $\om^{run}_L$ using 
\beq
\om^{run}_L\,\,=\,\,\bar \as(q^2_{av}= q q_0)\{\,
2.772\,\,-\,\,18.3\,\bar \as(q^2_0)\,\}
\eeq
instead of \eq{PAR}. This result was first obtained in Refs.
\cite{ORSAY}\cite{KM}. We will show below that the same scale should be  
incorporated  in the  calculation of $D$.

The last term in \eq{PSISP} is the most interesting one, because it shows
the violation of the Regge-BFKL asymptotics due to the running QCD
coupling. It was suggested by Kovchegov and Mueller in Ref. \cite{KM}
using quite a different method and, we hope, that our derivation
presented here is more
transparent. \eq{PSISP} shows explicitly that  Regge-BFKL asymptotics
is only valid in the limited range of $y - y_0$, namely, 
$D(\,\as\, \om_L\,  b\,)^2 (y -y_0)^3\,\,\ll\,\,1$ or $(y
-y_0)\,\,\ll\,\,\as^{-\frac{5}{3}}$ \cite{LREN}\cite{KM}.

The question arises can we trust the term proportional to  $(y - y_0)^3$.
Indeed, we have made  a lot of assumptions , in the derivation  as well as
 using  
the steepest descent method (SDM). Therefore, we have to check whether all
our
assumptions are selfconsistent.   They are:
\begin{eqnarray} 
& \left(\,\frac{\om_L}{D
\,r_0}\,\right)^{\frac{1}{3}}\,\times\,\frac{\De^S
\,r_0}{\om_L}\,\,\gg\,\,1\,\,\,\,\,\,asymptotic
\,\,\,of\,\,\,Airy\,\,\,function \label{ASSUMP} &\\
& \kappa^S\,=\,\frac{\de r \,\om_L}{r_0\,\De^S}\,\,\gg\,\,1\,\,\,\,\,
assumption\,\,\,\,used\,\,\,to\,\,obtain\,\,\De^S \label{ASSUMP1} &\\
& \frac{1}{3!}\,\frac{d^3 \Psi}{d^3 \De}\,|_{\De =
\De^S}\,\times\,\left(\,\frac{2!}{\frac{d^2 \Psi}{d^2 \De}\,|_{\De =
\De^S}}\,\right)^{\frac{3}{2}}\,\,\ll\,\,1\,\,\,\,\,\,selfconsistency\,\,of
\,\,the\,SDM  \label{ASSUMP2}
&\\
& 
(y\,-\,y_0)\,\,\leq\,\,\frac{r^2_0}{4
D}\,\,\,\,\,applicability\,\,of\,\,pQCD\,\,\,\,(\,see\,\,\,Ref.
\cite{KM}\,) \label{ASSUMP3}&
\end{eqnarray}
In addition, we have to check that $\frac{d^2 \Psi}{d^2 \De}|_{\De
=\De^S}$ is positive, but it is easy to find out that it is the case for
$D\,>\,0$. However, it is an indication that $D\,<\,0$ it has to be
considered separately. It turns out that the second equation of
\eq{ASSUMP} is the most restrictive . Taking $\De^S $ from \eq{SP3} we see
that it leads to
\beq \label{CHECK}
y \,- \,y_0\,\,\leq\,\,\left(\,\frac{\de
r\,r_0}{\om_L\,4\,D}\,\right)^{\frac{1}{2}}\,\,\propto\,\,\sqrt{\frac{\de
r}{\as^3}}
\eeq
Substituting \eq{CHECK} in $( y - y_o)^3$-term in $\Psi$ ( see \eq{PSISP})
we obtain the estimate on maximum value of this term which we can
guarantee in our approximation:
\beq \label{ESTIMATE}
\frac{D \om^2 _L}{r^2_0}\,(y -
y_0)^3\,\,\leq\,\,\sqrt{\frac{\om_L}{D\,r_0}}\,\times\,\left(\, \de
r\,\right)^{\frac{3}{2}}\,\,\propto\,\,\de r \,\sqrt{\frac{\de
r}{\as}}\,\,.
\eeq
One can see that in wide region $\de r \,>\,( \as)^{-\frac{1}{3}}$,  it is
legitimate to keep this term and, therefore, we can conclude, that
the running QCD coupling does not satisfy  the Regge-type asymptotics.
However, we
have to understand this result better since, at first sight,  zeros of
the denominator in \eq{SOL} yield  Regge asymptotics.

\subsection{High energy asymptotics ( a more detail analysis )}

{ $\mathbf D\,\,>\,\,0$}

First,  Airy functions have   zeros only at the negative values
of the argument,  and  their position   can be found with good
accuracy from the simple equation\cite{MATH}:
\beq \label{EQZE}
z\,\,=\,\,-  \, (\,\frac{3\pi n}{2} \,-\,\frac{3 \pi}{8}\,)^{\frac{2}{3}}
\,\,,
\eeq
where $z$ is the argument of the Airy function and $n$ is arbitrary
integral number ($n$ = 0,1,...).

Taking the argument of the Airy function in the denominator of \eq{GRNRF}
we obtain \cite{LREN}\cite{HKK}
\beq \label{ZEAI}
\De_n\,\,=\,\,-\,\om_L\,\,\left(\,\frac{D}{\om_L\,r^2_0}\,
\right)^{\frac{1}{3}}\,\times\,
 (\,\frac{3\pi n}{2} \,-\,\frac{3 \pi}{8}\,)^{\frac{2}{3}}\,\,.
\eeq
Here,$\De_n\,\,=\,\,\om_n\,\,-\,\,\om_L$ and one can see that all poles
in  $\om$ are located to the left from $\om_L$ ( $ \om_n\,\,<\,\,\om_L$
and $\De_n\,\,\propto \,\, \as^{\frac{2}{3}}\,\om_L$ ).  Therefore, we can
legitimately calculate $\De_n$ in framework of our approach. The whole
structure of singularities in the $\om$ - plane is  as follows
\cite{LREN}\cite{HKK}: 

1. the rightmost pole $\om_0$ is located to the left of $\om_L$ but,
theoretically, very close to it. Namely, its position is
\beq \label{OMO}
\om_0\,\,=\,\,\om_L\,\, 
-\,\om_L\,\,\left(\,\frac{D}{\om_L\,r^2_0}\,\right)^{\frac{1}{3}}\,\times\,
 (\,\frac{3 \pi}{8}\,)^{\frac{2}{3}}\,\,.
\eeq
 
2. for large $n \,\rightarrow \,\infty$  
\beq \label{LRGN}
\om_n\,|_{n\,\gg\,1}\,\rightarrow\,\,
\left(\,\frac{D}{r^2_0}\,\right)^{\frac{1}{3}}\,\times\,\frac{2}{3\pi
n}\,\,;
\eeq

3.  there fore, the solution has the infinite number of poles in
$\om$-plane
to the left of  $\om_0$ , which accumulate at $\om=0$ at
$n\,\rightarrow\,\infty$.

This picture of singularities, justifies  our saddle point calculation, 
since the position of the saddle point turns out to be shifted to
the right of $\om_L$. Note, that the contour of $\om$-integration is
chosen to be located to the right of all singularities of the integrand.

We can evaluate  the integral in a different way, namely, calculating each
pole
separately. In this case we have the asymptotic  form
\beq \label{POLES}
G_r(y - y_0,r,r_0)\,\,=\,\,\sum^{\infty}_{n=0}\,e^{\om_n \,(y\, -  \,y_0
)}
V(\om_n, r)\,\,,
\eeq
where $V( \om_n, r)$ is the $\om_n$ - pole.
This series was investigated in Ref.\cite{HKK} where it was shown that the
saddle point approximation effectively describes the sum over $n$ at
sufficiently large $n$. The right most singularity $\om_0$  gives a
suppressed contribution because of smallness of its residue, while $\om_n
\,\rightarrow\,0$
at large $n$,  but the residue is rather big.

{ $\mathbf D\,<\,0$ }

In this case the general solution has  the same form of \eq{GRNRF} but the
structure of singularities changes crucially: 

1. all poles $\om_n$ are located to the right of $\om_L$;

2. the position of the leftmost pole is
\beq \label{OMON}
\om_0\,\,=\,\,\om_L\,\,
+\,\om_L\,\,\left(\,\frac{D}{\om_L\,r^2_0}\,\right)^{\frac{1}{3}}\,\times\,
 (\,\frac{3 \pi}{8}\,)^{\frac{2}{3}}\,\,;
\eeq

3. at large $n$ $\om_n\,\,\rightarrow\,\,\infty$ accordingly the following
expression:
\beq \label{LOM}
\om_n\,|_{n \,\gg\,1}\,\,=\,\,\frac{|D|}{r^2_0}\,\times\,\left(\,\frac{3
\pi\,n}{2}\,\right)^2\,\,.
\eeq

In such a situation we cannot use the saddle point method and should
rather analyze \eq{POLES} with $\om_n$ given by \eq{LOM}.  
It is easy to see that \eq{POLES} can be reduced to the form:
\beq \label{POLESN}
G_r(y -
y_0,r,r_0)\,\,=\,\,\sum^{\infty}_{n\,\gg\,1}\,e^{\frac{|D|}{r^2_0}\,(\,\frac{3
\pi n}{2}\,)^2 \,(y -  y_0 )\,\,\pm\,\,i\,\pi \,n\,\frac{\de 
r}{r_0}}\,\,.
\eeq

To evaluate  this sum looks hopeless, at least, we do not see how it is
possible to do . However, we have  to go back to our derivation of
solution
(see \eq{3.1}), namely, to the integration over $f$
which led to the Airy function. The value of the typical $f$ ( $f^S$ ) in
this
integral can be evaluated using the saddle point approach and it is equal
to
\beq \label{FV}
f^S\,\,=\,\,\pm\,\sqrt{\frac{\om}{|D|}\,(\,r\,\,-\,\,r_0\,\frac{\om_L}{\om}\,)}
\,\,,
\eeq
which is of the order of $n$ for $\om_n$. Therefore, for large $n$ we
cannot trust the solution of \eq{GRNRF} and have to generalize the
solution
including the next term in expansion of $\om_{conf}(f)$ in \eq{OMD},
namely,
\beq \label{OMTT}
\om_{conf}\,\,=\,\,\om_L\,\,+\,\,D\,f^2\,\,-\,\,B\,f^4\,\,,
\eeq
where  $B$ was calculated in Ref.\cite{ROSS} 
\beq \label{B}
B\,\,=\,\,-\,\bar \as(q^2_0)\,\,\{\,64.294\,\,-\,\,\bar
\as(q^2_0)\,2756\,\}\,\,.
\eeq
One can see that $B\,\,>\,\,0$  everywhere,  except of  the region of very
small  $\as$.

One can see that \eq{OMTT} can be written as
\beq \label{QF}
\om_{conf}\,\,=\,\,\tilde{\om}_L\,\,-\,\,B \,(\, f^2 \,-\,f^2_0\,)^2 \,\,,
\eeq
where 
$$\tilde{\om}_L\,\,=\,\,\om_L \,\,+\,\,B\,f^4_0$$
 and
$$f^2_0\,\,=\,\,-\,\,\frac{|D|}{2 B}\,\,\,\,; \,\,\,\,\,\,\,\,\,
f_0\,\,=\,\,\pm \,i
\sqrt{\frac{|D|}{2 B}}$$.

Introducing a new variable $ f\,\,=\,\,f_0\,\,+\,\,\nu$ and integrating
in \eq{SOL} from $f_0$ we obtain 
\beq \label{GS1}
G(y,r)\,\,=\,\,
\eeq
$$
\int^{a + i\infty}_{a - i \infty} \,\frac{d \om}{
2 \,\pi\,i}\int^{\hat{f}  + i \infty}_{\hat{f}  - i \infty}\,\frac{d
\nu}{2\,\pi\,i}
\tilde{g}(\om)\,\,cos\left( \de r\,\,\sqrt{\frac{|D|}{2 B}}\,\right) \,e^{
\om \,(y - y_0) \,+\,\,\nu\,r
\,-\,\frac{r_0\,(\,\tilde{\om}_L \,\,\nu\,\,+
\,\,\frac{B\,f^2_0}{3}\,\nu^3\,\,)  
}{\om}}\,\,.
$$

 Integration over $\nu$ in \eq{GS1}  gives the Airy function
$Ai \left(\,(\frac{\om}{r_0 \,D'})^{\frac{1}{3}}\,[ r
\,- \,\frac{\tilde{\om}_L}{\om}\,r_0]\,\right)$  with 
\beq \label{DD}
D'\,\,=\,\,2\,|D|\,\,.
\eeq
Finally, the solution is very similar to \eq{GRNRF} with new $D'$ of
\eq{DD},namely

\beq \label{OTRGR}
G_r(y-y_0,r,r_0)\,\,=
\eeq
$$
\sqrt{\frac{r}{r_0}}\,\int^{a + i\infty}_{a -
i\infty}\,\frac{d \om}{2\pi i}\,\,cos\left(\de r \,\sqrt{\frac{|D|}{2 
B}}\right)\,\,e^{\om(y - y_0)}\,\,\frac{Ai \left(
(\frac{\om}{r_0
D'})^{\frac{1}{3}}\,[\,r\,-\,\frac{\om_L}{\om}r_0\,]\,\right)}{Ai\left(    
(\frac{\om}{r_0
D'})^{\frac{1}{3}}\,[\,r_0\,-\,\frac{\om_L}{\om}r_0\,]\,\right)}\,\,.
$$
Therefore, \eq{PSISP} gives the answer in this case,  for the saddle
point value of $\Psi$ with $D = D'$. The Green function $G_r(y - y_0,
r,r_0)$ has the form:
\beq \label{GRFROT}
G_r(y - y_0,r,r_0 )\,\,=\,\,\frac{1}{q \,q_0}\,\,\sqrt{\frac{\pi\,(\,\de
r\,)^2}{2\,D' ( y - y_0)^3}}\,cos\left( \de r\,\sqrt{\frac{|D|}{2
B}}\right)\,\,e^{\Psi(\De^S, D\,\rightarrow\,D')}\,\,.
\eeq
The unpleasant fact which is a direct consequence of the NLO BFKL is the
oscillation, which has first  been found in Ref. \cite{ROSS}.
This is a serious defect of the NLO BFKL, since $G_r$ is proportional to
the total cross section. Certainly, we cannot have faith in  such an
approach
which leads to  negative total cross section.

\subsection{Asymptotics at ultra high energies}

We can calculate the asymptotics of this Green function ( $G_r$ ) even at
very high values of energy.  Indeed,  \eq{ASSUMP3} which restricts the
value of $y - y_0$ does not affect $G_r$ since the nonperturbative
corrections have been absorbed in $G_{in} $ in \eq{SOLR}.  \eq{SOLR} 
gives a nice example how we can separate the unknown nonperturbative 
input distribution ( $ G_{in} $ ) and perturbative BFKL Green function.
Unfortunately, we are not able to do so in the case of the second Green
function ( $G_y$ ) for which \eq{ASSUMP3} is a restriction ( see Ref.
\cite{KM} for details ).

It is easy to see that at very large value of $y - y_0\,\,\gg\,\,\as^{-
\frac{5}{3}}$ the inequalities of  \eq{ASSUMP} - \eq{ASSUMP3} are violated 
and actually we cannot use the asymptotic expression for the Airy
functions both in numerator and in denominator. Therefore, the only source
of the asymptotics is the zero of the denominator at $\om \,\,=\,\,\om_0$,
where (see \eq{OMO} )
\beq \label{UA1}
\om_0\,\,=\,\,\om_l\,\,-\,\,\om_L\,\left(\,\frac{D}{\om_L\,r^2_0}\,\right)\,
\left(\,\frac{3\,\pi}{8}\,\right)^{\frac{2}{3}}\,\,\,.
\eeq
 Closing contour in $\om $ - integration on this pole we obtain
\beq \label{UA2}
G_r( y - y_0, r, r_0 )\,\,\,=
\eeq
$$
\frac{8}{\left(\,\frac{3\,\pi\,\om^2_L\,r^4_0}{8\,D^2}\,\right)^{\frac{1}{6}}}
\,\,Ai\left(\,(\frac{\om_L}{r_0\,D})^{\frac{1}{3}}\,\de r
\,\,-\,\,(\frac{3
\pi}{8})^{\frac{2}{3}}\,\right)\,\times\,e^{\om_0\,(\,y\,-\,y_0\,)}\,\,.
$$
Therefore, we obtain the typical Regge asymptotics without any  diffusion
in the log of transverse momentum. The first factor is the Reggeon residue
while the second one is the Reggeon propagator.

For the case $ D\,<\,0$  one can see that the asymptotics will be 
given by  \eq{UA1} with obvious substitute $
\om_L\,\rightarrow\,\tilde{\om}$ (see \eq{QF} ) and $
D\,\rightarrow\,D'\,=\,2\,|D|$ ( see \eq{DD} ).

One can see, that Regge behaviour depends only on initial virtualities
$Q_0 ( r_0 ) $ and corresponds to the pole in the angular momentum with
the position which is independent from  any characteristics of the ``hard"
processes. We hope, that this example can be instructive for high energy
phenomenology. At least, it gives an answer to the question:  why and 
how the Regge-like behaviour could appear for the initial condition for
the DGLAP evolution equations. Recall, that such a behaviour is heavily
used in the solution of the DGLAP evolution equations.
 
\section{$\mathbf G_y(y,r)$}
\setcounter{equation}{0}
\subsection{A general solution}
 To find $G_y (y, r )$ we return to a general equation of \eq{RUNBFKL}.
 The inhomogeneous term in it corresponds to \eq{BFKL} or, in other words,
the inhomogeneous \eq{RUNBFKL} is written for  $G_y (y, r )$.
One can easily  find   the solution to this equation, taking the
solution of
the homogeneous equation ( see \eq{SOL} ) but considering $\tilde{g}$ as
a function of $\om$ and $f$. Substituting it  back to \eq{RUNBFKL} we
obtain the following equation for $\tilde{g}(\om,f)$:
\beq \label{GY1}
-\,\,\om\,\frac{d \tilde{g}(\om, f)}{d f}\,\,\,=\,\,\,r_0 \,e^{-f\,r_0}\,
e^{\frac{r_0}{\om}\,\int^f_{f_0})\,\om_{conf}(f')\,d \,f'}\,\,.
\eeq
Finally, a general solution can be reduced to the form:
\beq \label{GY2}
G_y(y - y_0, r, r_0)\,\,=\,\,
\int^{a + i\infty}_{a - i \infty} \,\frac{d \om}{
2 \,\pi\,i}\int^{\hat{f} + i \infty}_{\hat{f} - i \infty}\,\frac{d
f}{2\,\pi\,i}\,\,
\frac{r_0}{\om}\,
\,\,\int^{\infty}_{f}\,\frac{d f'}{2 \pi \,i}
\eeq
$$
\,e^{ \om \,(y - y_0)
\,+\,f\,r\,\,-\,\,f'\,r_0 
\,-\,\frac{r_0}{\om}\,\int^{f'}_{f}\om_{conf}(f'')\,d \,f''}\,\,.
$$
The integral over $\om $ can be evaluated  and we obtain  the general
solution in
the form that we will use in this paper;
\beq \label{GY3}
G_y(y - y_0, r, r_0)\,\,=\,\,r_0\,\int\int \,\frac{d f \,\,d
f'}{2\,\pi\,i}\,\theta( f' \,-\,f )\,e^{f\,r \,-\,f' r_0}\,\,
\eeq
$$
I_0 \left(
\,2\,\sqrt{(y - y_0 )
\,r_0
\,\int^{f'}_{f}\,\om_{conf}(f'')\,d f''}\,\right)\,\,.
$$
It is easy to check that $G_y(y - y_0, r, r_0)$, given by \eq{GY3},
approaches $\de( r - r_o)$ at $y\,\rightarrow\,y_0$.

Since at $y - y_0 \,\gg\,1$ the argument of $I_0$ is large,  we can use
the
asymptotics of the modified Bessel function, namely, $I_0(z)\,|_{z
\,\gg\,1}\,\rightarrow\,\frac{1}{\sqrt{2 \pi z}}\,e^{z}$. To get a more
compact answer we introduce: (i) new variables $ f^{-}\,=\,f' - f $ and
$f^{+}\,=\,f' + f $; and  (ii) the integral representation for $\theta$ -
function
\beq \label{THETA}
\theta ( f^{-} )\,\,=\,\,\int^{\mu_0 \,+\,i\,\infty}_{\mu_0\,-\,i\,\infty}  
 \,\,\frac{d \mu}{2 \,\pi \,i\,\mu}\,e^{\mu f^{-}}\,\,.
\eeq
The final result is
\beq \label{GY4}
G_y(y - y_0, r, r_0)\,\,=\,\,r_0\,\int \int \int  \,\frac{d f \,\,d
f'\,d \,\mu }{( 2\,\pi\,i )^2\,\mu }
\eeq
$$
e^{ -\,(\,\frac{r + r_0}{2} -
\mu\,\,)\,f^{-} \,\,+\,\,\frac{ r - r_0}{2}\,f^{+}\,\,+\,
\,2\,\sqrt{(y - y_0 )
\,r_0
\,\int^{\frac{f^{+}\,+\,f^{-}}{2}}_{\frac{f^{+}\,-\,f^{-}}{2}}\,
\om_{conf}(f'')\,df''}}\,\,.
$$
\subsection{Regge-BFKL  asymptotics}
   
We assume that both  $f^{-}$ and $f^{+}$ are small enough to use \eq{OMD}
for $\om_{conf}(f)$. We can see that the answer has a form:
\beq \label{GYPSI}
G_y(y - y_0, r,r_0)\,\,=\,\,\,\int \int \int \frac{d\,
f^{+}\,\,d\,f^{-}\,\,d\,\mu}{ (\, 2\, \pi\, i\,)^2\,\,\,\, 
\mu}\,\,e^{\Psi(y,r,f^{+},f^{-}
)}\,\,,
\eeq 
and  taking
explicit  integration over $f''$,  $\Psi$ can be written as
\begin{eqnarray} \label{PSIY}
&
\Psi(y,r,f^{+},f^{-})\,\,=\,\,-\,(\,\frac{r + r_0}{2} -
\mu\,)\,f^{-} \,\,+\,\,\frac{ r -
r_0}{2}\,f^{+}
+& \\
&
\,\,2\sqrt{\om_L \,r_0\,( y \,-\,y_0)}\,\
\sqrt{f^{-}}\,\cdot\,\{ 1 \,\,+\,\,\frac{
D}{6 \,\om_L}\,[\,\frac{3}{4}\,( f^{+})^2\,\,+\,\,\frac{1}{4}\,( f^{-}
)^2\,]\,\}\,\,.& \nonumber   
\end{eqnarray}
We evaluate  the integrals over $f^{-}$ and $f^{+}$,  using the steepest
descent
method in which we have the following equations for the saddle points :
\beq \label{SPGY}
\frac{\partial \Psi}{\partial f^{-}}\,|_{f^{-} = f^{-,S}}\,\,
=\,\,0\,\,;\,\,\,\,\,\,\, 
\frac{\partial \Psi}{\partial f^{+}}\,|_{f^{+} = f^{+,S}}\,\,=\,\,0\,\,.
\eeq

\eq{SPGY} gives
\begin{eqnarray}
&
\frac{ r \,+\,r_0}{2} \,\,-\,\,\mu\,\,\,+
\,\,\frac{\sqrt{\om_L\,r_0\,(
y\,-\,y_0)}}{\sqrt{f^{-,S}}}\,\,\times\,\,\{\,1
\,\,+\,\,\frac{D}{8\,\om_L}\,[ \,( f^{+,S} )^2\,+\,\frac{5}{3}\,( f^{-,S}
)^2
\,]\,\}\,\,=\,\,0
&\label{SPY1} \\
&
\frac{r\,-\,r_0}{2}\,\,+\,\,\sqrt{\om_L\,r_0\,( y  
\,-\,y_0)}\,\sqrt{f^{-}}\,\,\times\,\,\frac{D}{2
\om_L}\,f^{+,S}\,\,=\,\,0\,\,.
 \label{SPY2}&
\end{eqnarray}
The leading order solutions of the above equations are
\begin{eqnarray} 
&
f^{-,S}_0\,\,=\,\,\left(\,\frac{\sqrt{\om_L\,r_0\,( y\,-\,y_0)}}{\frac{r +
r_0}{2} \,-\,\mu}\,\right)^2\,\,;
&\label{SPY3}\\
&
f^{+,S}_{0}\,\,=\,\,- \,\frac{\frac{r \,-\,r_0}{2}}{\frac{D}{2
\om_L}\,\,\sqrt{\om_L\,r_0\,( y\,-\,y_0)}\,\,\sqrt{f^{-,S}_0}}
\,\,. \label{SPY4}&
\end{eqnarray}
Keeping the next order term we have
\begin{eqnarray}
&
f^{-,S}\,\,=\,\,\left(\,\,\frac{\sqrt{\om_L\,r_0\,( y\,-\,y_0)}}{\frac{r
+ r_0}{2} \,-\,\mu}\,\right)^2\,\times\,\{\,1\,\,+\,\,\frac{D}{4
\om_L}\,[ (f^{+,S}_0)^2\,\,+\,\,\frac{5}{3}\,( f^{-,S}_0 )^2\,]\,\};
& \label{SPY5}\\
&
f^{+,S}\,\,=\,\,-\,\frac{\frac{r \,-\,r_0}{2}}{\frac{D}{2 
\om_L}\,\,\sqrt{\om_L\,r_0\,(
y\,-\,y_0)}\,\,\sqrt{f^{-,S}_0}}\,\{\,1\,\,-\,\,\frac{D}{8 
\om_L}\,[ (f^{+,S}_0)^2\,\,+\,\,\frac{5}{3}\,( f^{-,S}_0 )^2\,]\,\}
\,\,. \label{SPY6}&
\end{eqnarray}
Substituting \eq{SPY5} and \eq{SPY6} in \eq{PSIY}, we obtain for the
saddle value of $\Psi$
\begin{eqnarray} 
&
\Psi\left(y,r,f^{+,S},f^{-,S}\right)\,\,=\,\,\om_L\,\frac{2\,r_0}{ r_0
\,+\,r}\,(\,y\,-\,y_0\,)\,\,-\,\,\frac{ ( \de r )^2}{4\,D\,\frac{2
\,r_0}{r_0\,+\,r}\,(y\,-\,y_0)}\,\,+
& \nonumber\\
&
\,\,\frac{1}{12}\,\,D\,\frac{r +
r_0}{2
r_0}\,\,\left(\,\om_L\,\frac{2 r_0}{r_0 + r}\,\frac{r_0 + r}{2}\,\right)^2
\,\,(  \,y \,-\,y_0\,)^3\,\,. \label{PSISPY}
&
\end{eqnarray}
Here, we have taken the integration over $\mu$. Indeed, it is easy o see
that the saddle point value of
$\Psi\left(\,y,r,f^{+,S},f^{-,S} \right)\,\,\rightarrow\,\,-\,\infty$ when
$\mu \,\,\rightarrow\,\,- \infty$. Therefore, we can close contour of
integration over $\mu$ and take a residue at $\mu = 0$.

In \eq{PSISPY} one can see the power increase (Reggeon - like) with energy
( the first term ) and BFKL diffusion in log of transverse momenta ( the
second term). It is interesting to note that in both terms as well as
in the third one, the running QCD
coupling constant depends on the scale $q^2_{ev} \,=\,q\,q_0 ( r_{ev}
\,=\,\frac{r_0 + r}{2} )$.  Namely, coupling constant at this scale enters
in calculation of  both $\om_L$ and $D$.  

For both Green functions we obtain  a term proportional to $( y - y_0)^3$
which  was first found in Ref. \cite{KM} \footnote{The numerical
coefficient in front of this term ($\frac{1}{12}$)  is different from the
coefficient calculated in Ref. \cite{KM} ($\frac{1}{3}$) due to
different definition of $D$ ( $D ( our )$ = 4 $ D( Ref.  \cite{KM})$ ).}.

\subsection{Regge-BFKL  asymptotics for $\mathbf D\,<\,0 $}
One can see from \eq{SPY3} that the saddle point value of $f^{-}$ does not
depend on the value of $D$. It suggests the following logic of approach:
we expand $\Psi$ in   a  general solution of \eq{GY4} in the region of
small
$f^{-}$ while  keeping $f^{+}$ to be arbitrary large. In such an
approximation we have
$$
\frac{d \int^{\frac{f^{+} + f^{-}}{2}}_{\frac{f^{+} -
f^{-}}{2}}\,\,\om_{conf}(f') \,d\,f'}{d
f^{-}}\,\,=
\,\,
\frac{1}{2}\,\{\,\om_{conf} \left(
 \frac{f^{+} + f^{-}}{2}
\right)\,\,+\,\,\om_{conf} \left( \frac{f^{+} - f^{-}}{2}  
\right)\,\}\,\,=
$$
\beq \label{OTD1}
\,\, \om_{conf} \left( \frac{f^{+}}{2}
\right)\,\,+\,\,\frac{1}{8}\,\,{\om''}_{conf}\left( \frac{f^{+}}{2}
\right)\,( f^{-} )^2\,\,,
\eeq
where ${\om''}_{conf}\,\,=\,\,\frac{d^2 \om_{conf} (f)}{d^2 f}$.

The value of $f^{-}$ in the saddle point is equal to
\beq \label{SPOTD}
\sqrt{f^{-,S}}\,\,=
\eeq
$$
\frac{2}{r_0 + r}\,\sqrt{ r_0
\om_{conf}\left(\frac{f^{+}}{2} \right)\,( y - y_0  
)}\,\,\left(\,1\,\,+\,\,
\frac{5}{48}\,\frac{{\om''}_{conf}\left(
\frac{f^{+}}{2}\right)}{\om_{conf}\left(\frac{f^{+}}{2}
\right)}\,\,(\,f^{-,S}_0)^2\,\,\right)\,\,,
$$
where
\beq \label{SPOTD1}
\sqrt{f^{-,S}_0}\,\,=\,\,\frac{2}{r_0 + r}\,\sqrt{ r_0
\om_{conf}\left(\frac{f^{+}}{2} \right)\,( y - y_0)}\,\,.
 \eeq 

$\Psi$ for  this saddle point is equal to
\beq \label{PSI0}
\Psi \left( y - y_0,r, f^{+},f^{-,S} \right)\,\,=
\eeq
$$
\frac{\de r}{2}\,\,+\,\,\frac{2 r_0}{r_0
\,+\,r}\,\om_{conf}\left(\frac{f^{+}}{2}
\right)\,(\,y\,-\,y_0\,)\,\lbrack\,
1\,\,\,+\,\,\frac{1}{24 }\,\,\frac{{\om''}_{conf}\left(
\frac{f^{+}}{2}\right)}{\om_{conf}\left(\frac{f^{+}}{2}
\right)}\,\,(\,f^{-,S}_0)^2\,\,\rbrack\,\,.
$$
The position of the saddle point for $f^{+}$ integration is located near
to the minimum of $\om_{conf}( \frac{f^{+}}{2} )$.  If we denote the
position of minimum $f_0\,\,=\,\,Re f_0\,\,+\,\,i Im f_0 $ we can obtain
the result after taking the integral over $f^{+}$ using the saddle point
method. In practice as we have mentioned in the NLO BFKL the minimum 
occurs at $f_0\,\,=\,\,\pm\,i\, \sqrt{\frac{|D|}{2 B}}$ ( see \eq{QF} )
for $D \,<\,0$ and $f_0 = 0$ for positive $D$. 

The result is
\beq \label{GENR}
G_y ( y - y_0,r,r_0)\,\,=
\eeq
$$
\frac{1}{q\,q_0}
\,\sqrt{\frac{\om_{conf}\left( \frac{f_0}{2}
\right)}{{\om''}_{conf}\left(\frac{f_0}{2} \right)( \frac{r_0 + r}{2})^3}}
\,\,\times\, cos\left( 2 Im f_0 \de r \right)
 \,\times\,e^{\Psi \left(y -
y_0,r,r_0,f^{+,S},f^{-,S} \right)}\,\,,
$$
where
\beq \label{PSIFIN}
\Psi \left(y - y_0,r,r_0,f^{+,S},f^{-,S}
\right)\,\,=
\frac{\om_{conf}\left( \frac{f_0}{2} \right)\,2\,r_0}{r_0
+   r}\,\,(\,y\,-\,y_0\, )
\eeq
$$
-\,\,\frac{( \de r)^2}{2
{\om''}_{conf}\left(\frac{f_0}{2}
\right)\,(\,y\,-\,y_0\,)}\,\,+\,\,\frac{1}{24}\,
{\om''}_{conf}\left(\frac{f_0}{2}\right)\,\frac{\om^2_{conf}\left(
\frac{f_0}{2} \right)}{ (\frac{r_0 + r}{2})^2}\,(\,y\,-\,y_0 )^3\,\,.
$$

\section{Summary}
\setcounter{equation}{0}
In this paper we analyze the prediction for high energy asymptotics that
emerges from the NLO BFKL equation. Below  we  summarize the
results of our analysis.

1. We found two Green functions ( see \eq{GRNRF} and \eq{GY4} )that govern
the asymptotic behaviour
of the scattering amplitudes at high energy with two different initial
conditions: at fixed virtuality $Q^2 \,=\,Q^2_0 $ and at fixed
energy ($x$).

2. For sufficiently small values of energy \cite{LREN}\cite{KM}
$$ y\,\,=\,\,\ln(s/qq_0)\,\,\leq\,\,\as^{ - \,\frac{5}{3}}$$
in both Green function we found the Regge - BFKL asymptotics
\beq \label{5.1}
G\,\,\propto\,\,e^{\om_a\,(\,y\,-\,y_0\,)\,\,-\,\,\frac{(\,r\, -\, r_0\,
)^2}{4\,D_a\,
(\,y\,\,-\,\,y_0\,)}}\,\
\eeq
However,  parameters $\om_a$ and $D_a$ in \eq{5.1} turns out to be
different for different sign of $D$ in the NLO expression for the BFKL
kernel (see \eq{OMD})
\parbox{6.5cm}{
\begin{flushleft}
$\mathbf D\,\,>\,\,0$ \\
$\om_a\,\,=\,\,\frac{2 r_0}{r + r_0}\,\om_L$\,\,\,\,(see\,\,\eq{OMD})\\
$D_a\,\,=\,\frac{2 r_0}{r + r_0}\,\,D$
\end{flushleft}
}
\parbox{9.5cm}{
\begin{flushleft}
$\mathbf D\,\,<\,\,0$ \\
$\om_a\,\,=\,\,\frac{2 r_0}{r + r_0}\,\,\lbrack\,\om_L\,\,+
\,\,\frac{D^2}{4 B}\,\rbrack$ \,\,\,\,(see\,\,\eq{QF})\\
$D_a\,\,=\,\,2\,\frac{2 r_0}{r + r_0}\,|D|$
\end{flushleft}}

3.  We confirm the result of Ref. \cite{KM} that the first corrections to
the Regge-BFKL asymptotics  lead to  an  additional term in the
exponent of \eq{5.1}, which has a form:
\beq \label{5.2}
\frac{1}{12}\,\frac{\,D_a\,\,\om^2_a}{r^2_0}\,\,(\,y\,\,-\,\,y_0\,)^3\,\,.
\eeq
We showed that it is legitimate to consider such a term in the framework
of our approach.

4. Unfortunately, we found the oscillation in $r - r_0$ for the asymptotics
in the NLO BFKL in the huge region of initial virtualities $q^2_0 ( r_0)$
where the NLO generates $D\,<\,0$  ( see \eq{QF} ). This
oscillation was first suggested in Ref. \cite{ROSS}. Indeed, the
asymptotics has a factor ( see \eq{GENR} )
\beq \label{5.3}
G_y\,\,\,\propto\,\,\,cos \left(\,2\,\,(\,r \,-\,r_0\,)\,\,\sqrt{\frac{| D
|}{2 B}}\,\right)\,\,.
\eeq 
We consider this result as a challenge for the experts in QCD since
it is  a strong indication that the NLO BFKL approach
has a pathology. We do not think that an  additional integration in
Eq.(2.1) will rule out this oscillations.It should be stressed that the
exchange of the BFKL Pomeron has a clear physical meaning as the
asymptotics of the colour dipole - dipole scattering ( see Ref.\cite{MU94}
). It is difficult to believe that anyone would cope with a  negative
total
cross section for such a  process \footnote{ It should be stressed that
it was first demonstrated in Ref.\cite{BV} that the NLO BFKL corrections
to the anomalous dimension lead to the negative probability at
exceedingly small $x$. Perhaps, this result is correlated with ours, but
the difference is that we observe our oscillations in the region of
applicability of the BFKL approach, and, therefore, we cannot follow the
``wise" advise of Dokshitser \cite{DOK}:``Let DIS structure functions in
peace!"}.

5.  Our solution of \eq{GRNRF} suggests the asymptotic behaviour in the
region of large values of $Q^2$. Indeed, the Regge-BFKL asymptotics as
well as the term which violates it come from the saddle point
approximation in the region of $\om\,\,\approx\,\,\om_a$, but we find 
that $\om_a$ falls at $r \,\,\gg\,\,1$,  while the contribution from the
zeros of the denominator in \eq{GRNRF} do not depend on virtuality $Q^2 (
r ) $. Therefore, at large $Q^2$ the rightmost singularity in  the $\om$
integration will be the position of the zero 
\beq \label{5.4}
\om_0\,\,=\,\,\om_a
\,\,-\,\,\om_a\,\,\left(\,\frac{D_a}{\om_a\,\,r^2_0}\,\right)\,\,
\left(\,\frac{3\,\pi}{8}\,\right)^{\frac{2}{3}}\,\,.
\eeq
This singularity leads to Regge - behaviour 
$$
G_r\,\,\,\propto\,\,\,e^{\om_0\,(\,y\,\,-\,\,y_0\,)}
$$
in spite of the fact that in the huge region of intermediate values of
$Q^2$ we have a much more complicated behaviour. 

This observation can be used as a justification of the Reggeon-like 
behaviour of the structure functions in the region of small $x$, used for  
 the
initial condition for the evolution equations.

6. We cannot trust the perturbative QCD calculation  for the Green
function $G_y$ since at $ y \,-\,y_0\,\,\leq\,\,\frac{r^2_0}{4\,D}$
 ( see Ref. \cite{KM} ) the contamination from the nonperturbative QCD
region is rather big. We think, this is a principal difference between
$G_r$ and $G_y$ , that in $G_r$  all nonperturbative corrections can be
included in the input distribution $G_{in}$ in \eq{SOLR} while $G_r$ can
be calculated in the framework of perturbative QCD.

 \section*{Acknowledgements}
I wish to  acknowledge  very helpful discussions with 
 V. Fadin, L. Lipatov and A. Mueller on the broad subject of low $x$
physics, including the NLO BFKL equation. I  also thank Yu. Kovchegov for
stimulating and helpful discussions of his paper \cite{KM} before
publication  during the Minnesota Conference (April 1998). I am  also very
grateful to organizers of the Minnesota Conference for creating a 
working atmosphere  which reminds me of  the atmosphere of my youth.
My special thanks go to E. Gotsman and U. Maor without whose encouraging
optimism and every day support this paper will not be able to appear.

 \end{document}